\documentclass[showpacs,twocolumn,amsmath,superscriptaddress,pre]{revtex4-1}
\usepackage[dvips]{graphicx}
\usepackage{epsfig}
\usepackage{color}
\usepackage{subfigure}

\begin{document} 
\title{
Finite-size scaling of survival probability in branching processes
} 
\author{Rosalba Garcia-Millan}
\affiliation{Departament de Matem\`atiques,
Facultat de Ci\`encies,
Universitat Aut\`onoma de Barcelona,
E-08193 Barcelona, Spain}
\affiliation{Departament de F\'{\i}sica, 
Facultat de Ci\`encies,
Universitat Aut\`onoma de Barcelona,
E-08193 Barcelona, Spain}
\author{Francesc Font-Clos}
\affiliation{Centre de Recerca Matem\`atica,
Edifici C, Campus Bellaterra,
E-08193 Barcelona, Spain.
} 
\affiliation{Departament de Matem\`atiques,
Facultat de Ci\`encies,
Universitat Aut\`onoma de Barcelona,
E-08193 Barcelona, Spain}
\author{\'Alvaro Corral}
%
\affiliation{Centre de Recerca Matem\`atica,
Edifici C, Campus Bellaterra,
E-08193 Barcelona, Spain.
} 
\affiliation{Departament de Matem\`atiques,
Facultat de Ci\`encies,
Universitat Aut\`onoma de Barcelona,
E-08193 Barcelona, Spain}
\begin{abstract} 
Branching processes pervade many models in statistical physics. 
We investigate the survival probability of a Galton-Watson branching process after a finite number of generations. 
We reveal the finite-size scaling
law of the survival probability for a given branching process ruled by a probability distribution of the number of offspring per element whose standard deviation is finite,
obtaining the exact scaling function as well as the critical exponents.
Our findings prove the universal behavior of branching processes concerning the
survival probability.
\end{abstract} 

\maketitle

\section{Introduction}

Branching processes have become a very useful modeling tool 
\cite{Harris_original,Athreya},
originally in demography and population biology \cite{Kendall75},  
later in genetics 
and in the theory of nuclear reactions \cite{branching_biology,neutron_fluctuations},
and more recently in seismology \cite{Vere_Jones76,Helmstetter_Sornette_jgr02}.
Important applications in statistical physics arise due to the relationships
of branching processes with critical phenomena, 
through percolation theory and self-organized criticality (SOC) \cite{Christensen_Moloney,Zapperi_branching,Hergarten12}.

Indeed, branching processes provide one of the simplest examples of a second-order 
phase transition, equivalent to percolation in the Bethe lattice, 
and therefore free of geometric complexity \cite{Christensen_Moloney}.
These phase transitions (also called continuous)  
are characterized by a sudden change of an ``order parameter''
from zero to a non-zero value,
at precisely a critical value of  a ``control parameter'' \cite{Stanley,Yeomans1992}.
This fact is exploded in SOC theory,
with mean-field sand-pile models where avalanches propagate
through a system by means of a ``critical branching process'';
the peculiarity of SOC is that the critical state is reached in a spontaneous, 
self-organized way \cite{Christensen_Moloney,Jensen,Pruessner_book,Corral_FontClos}.

The common fact of criticality 
is the fulfillment of 
scaling laws for the thermodynamic variables
and the correlation length $\xi$
close to the critical point; e.g., 
in a magnetic system \cite{Stanley}, 
$$
\frac m {|t|^\beta}= g_{\pm} \left(\frac h {|t|^\Delta}\right),
$$
and an analogous one for $\xi |t|^\nu$,
where $m$ is the dimensionless magnetic moment per particle, 
which plays the role of order parameter,
$t$ is the reduced temperature, $h$ is the reduced magnetic field,
$\beta$, $\Delta$, and $\nu$ are critical exponents, 
and $g_\pm$ represents two scaling functions, 
one ($+$) for $t>0$ and another one $(-)$ for $t<0$.
The critical point is achieved at $t=h=0$.

A fundamental approach to analyze critical phenomena is by means
of finite-size scaling. It turns out to be that the sharp change 
in the properties of a system at a critical state is only possible in the
thermodynamic limit (this is the limit of infinite system size).
However, in practice, computer simulations cannot attain such limit, 
for obvious reasons, and one cannot infer the existence of 
a critical point from the results of computer simulations alone. 
Then, in a finite system of size $L$ an additional dependence appears, 
given by the ratio $\xi /L$, which can be replaced by
$L|t|^\nu$, yielding the ansatz
\begin{equation}\label{ffs_original}
m = |t|^\beta \tilde g_\pm
\left(
\frac h {|t|^\Delta}, L |t|^\nu
\right)
=\frac{1}{L^{\beta/\nu}} \mathcal{G}
\left(
a L^{\Delta /\nu} h, b L^{1/\nu} t
\right)
,\end{equation}
where $\tilde g_\pm$ and $\mathcal{G}$ are bivariate scaling functions 
and $a$ and $b$ are metric factors introduced to ensure
universality \cite{Privman}.  
The previous scaling law
is known as finite-size scaling. 
Note that, in the case when $t$ and $h$ appear linearly in the scaling function,
no distinction is made between $t>0$ and $t<0$ and a single
scaling function $\mathcal{G}$ is enough for describing both regimes.
The reason is that in a finite system there is no singularity
at $t=h=0$, where $\mathcal{G}$ is smooth and analytic \cite{Privman}.

In this paper we show that branching processes with size limitations
display finite-size scaling in the same way as in critical phenomena.
We are able to derive the exact form of the scaling function
as well as the critical exponents.
In the next section we review the basic language for branching processes
whereas in the third one we apply it to finite-size branching processes.
In the last section some implications are discussed.

Usually, when the distance to the critical point is kept fixed,
the decay of the probability of surviving
towards the infinite-size case is exponential in $L$
\cite{Harris_original,Athreya}.
In contrast, the finite-size scaling approach demonstrates,
keeping fixed the distance to the critical point in relative units of $1/L$, 
a power-law decay with $L$, resulting 
in a sort of law of corresponding states which turns out to be valid 
for any system size $L$ (provided that this is large enough) and
for any branching process of the Galton-Watson type (provided a finite variance).
This universality \cite{Yeomans1992,Stanley_rmp} arises because, as we show, 
at the critical point the only relevant quantity is the variance of the
distribution that defines the Galton-Watson process.

\section{Overview of previous important results}

For the connection of branching processes with critical phenomena
it is enough to consider simply the Galton-Watson process \cite{Harris_original}.
This is started, in the zeroth generation
of the process, by one single element,
which produces other elements, called offspring, 
in a number given by a random variable $K$.
The offsprings of the initial element constitute the first generation of the process,
and each one of them produces again a random number of offspring,
which are the second generation, and so on.
The main ingredient of the model is that the number
of offsprings of any element follows the same distribution (that of $K$),
and each of these random numbers is independent from those
of the other elements.

The first variable of interest is $N_t$,
which represents the number of elements in each
generation $t$. The initial condition is written then 
as $N_0=1$.
The key question in branching process is if the process gets
extinct or not, and this is represented by the event $N_t=0$;
in particular, as extinction is an absorbing state, all extinction
events are included in $N_t=0$ with $t\rightarrow \infty$.
Then, the probabilty of extinction is 
$$ P_\mathrm{ext}=\lim_{t\rightarrow \infty} \mbox{Prob}(N_t=0).$$ 

At this point it is very useful to introduce the probability generating function. 
Consider a generic discrete random variable $X$ which takes value 0 with probability $p_0$, value 1 with $p_1$, ... and value $x$ with probability $p_x$.
The probability generating function $f_X(z)$ is defined as
$$
f_X(z) = \sum_{x=0}^\infty p_x z^x =\langle z^X\rangle,
$$
where the dependence of $f_X(z)$ is on $z$, 
and $X$ indicates to which random variable it corresponds
(in our case, $X=K$ or $X=N_t$).
Useful but straightforward properties of $f_X(z)$ are the following:
(i) $f_X(0)=p_0$;
(ii) $f_X(1)=1$;
(iii) $f'_X(1)=\langle X \rangle$;
(iv) $f''_X(1)=\langle X (X-1)\rangle$;
(v) $f'_X(z) \ge 0$ in $[0,1]$;
(vi) $f''_X(z) \ge 0$ in $[0,1]$;
where the primes denote differentiation (with respect to the variable $z$),
and we assume those expected values exist and are finite.

Applying the first of these properties to the variable $N_t$, we get, 
for the probability of extinction,
$$ P_\mathrm{ext}=\lim_{t\rightarrow \infty} f_{N_t}(0),$$
which, as we will see, constitutes a great simplification of the calculation.
Using that $N_{t+1}=\sum_{i=1}^{N_t} K_{t,i}$, where $K_{t,i}$
is the number of offsprings of the $i-$th element in the $t-$th generation,
and that $f_{N_1}(z)=f_K(z)$,
it is possible to derive a fundamental theorem in branching processes, 
which is
$$
f_{N_t} (z) =f_K(f_K( \dots f_K(z) \dots))=f_K^t(z),
$$
where the superindex $t$ means composition $t$ times
(not power). 
This is valid because the $K_{t,i}$
are independent and identically distributed \cite{Harris_original,Corral_FontClos}.
Therefore, the probability of extinction turns out to be
$$ P_\mathrm{ext}=\lim_{t\rightarrow \infty} f_K^t(0),$$
i.e., the repeated iteration of the origin $z=0$ through the function
$f_K(z)$.

Using the rest of properties of probability generating functions listed above it
is possible to show that the probability of extinction is given by
$P_\mathrm{ext}=q$, where $q$ is the smallest non-negative fixed point of $f(z)$, 
i.e., we have $f_K(q)=q$. 
For $\langle K \rangle \le 1$ it turns out to be that $q=1$
but for $\langle K \rangle > 1$ one gets $0 \le q < 1$
\cite{Harris_original,Athreya,Corral_FontClos}.
As $q$ varies continuously with $\langle K \rangle$ this
reveals the existence of a continuous (or second order) phase
transition in the system, with control parameter $\langle K \rangle$
and with the probability of surviving to extinction,
or survival probability,
$P_\mathrm{surv}=1-P_\mathrm{ext}=1-q$ behaving as an order parameter.
This is zero below and at the critical point $\langle K \rangle =1$,
and strictly positive above the critical point, when
$q$ and therefore $P_\mathrm{surv}$ will depend on the parameters of the distribution of $K$.
The phase diagram of the Galton-Watson process
consists then of three regimes: subcritical, critical, and supercritical, 
depending on the value of $\langle K \rangle$.

\section{Finite-size effects}

In contrast with the explained above, we consider here
a system with a limitation in size, 
that is, the limit of infinite generations $t\rightarrow \infty$
cannot be reached and one has instead an imposed maximum
number of generations $L$, which we identify with system size.
So, we redefine extinction as extinction at the boundary $t=L$,
then, $P_\mathrm{ext}(L)=\mbox{Prob}(N_L=0)$.
One can see that this value will be smaller than in an infinite equivalent
system, as extinction at $t=L$ is a particular case of extinction at
$t \rightarrow \infty$.
In any case, as in the infinite system, we will have that
the probability is given by the iteration of the origin through $f(z)=f_K(z)$
(from now on, to ease the notation we drop the subindex $K$ in $f_K(z)$),
i.e.,
$$
P_\mathrm{ext}(L)=f_{N_L}(0)=f^L(0).
$$
So, the fixed point $q$ will not be reached but instead 
we will have that $P_{ext }(L)< q$.

What we expect is that after a large number of generations $n$
(if $L$ is also large), $f^n(0)$ will be close to the fixed point $q$,
i.e., $q-f^n(0)$ will be close to zero.
Then we will perform a Taylor expansion of $f(f^{n}(0))$, 
considered a function of $f^{n}(0)$, around the
abscissa point $q$, that is,
\begin{alignat*}{1}
&f^{n+1}(0)=f(f^{n}(0))=\\
\notag&=f(q)+f'(q)(f^{n}(0)-q)+\frac 1 2 f''(q)(f^{n}(0)-q)^2+\ldots,
\end{alignat*}
up to second order.
Using the fixed-point condition we can rewrite
$$q-f^{n+1}(0)=f'(q)(q-f^{n}(0))-\frac 1 2 f''(q)(q-f^{n}(0))^2;$$
in other words, the distance $d$ to the fixed point $q$ at iteration 
$n+1$ is given by
$$
d_{n+1}=M d_{n}-C d_{n}^2
$$
when $n$ is large, defining 
$M=f'(q) \mbox{ and } C=f''(q)/2$.
Actually, it is simpler to iterate the inverse of the distance, 
which at the lowest orders in $d_{n}$ is
$$
\frac 1 {d_{n+1}} = \frac 1 {M d_{n}}\left(1 +\frac {C d_{n}}{M}\right),
$$
and introducing the inverse of the distance, $c_n=1/d_n$,
$$
 {c_{n+1}} = \frac {c_{n}} {M} +\frac {C }{M^2}.
$$
It is easy to see that successive iterations lead to
\begin{alignat}{1}
\notag
 {c_{n+\ell}} &= \frac {c_{n}} {M^\ell} +\frac {C }{M^2}
\left(1 +\frac 1 M +\dots +\frac 1 {M^{\ell-1}}\right)=\\
&=\frac {c_{n}} {M^\ell} +  \frac{C(1-M^\ell)}{M^{\ell+1}(1-M)},
\label{dearriba}
\end{alignat}
using the formula of the geometric progression.
From here one can see that, for fixed $M$ and large $\ell$,
the decay of the distance to the fixed point $q$ is 
proportional to $M^{\ell}$, and therefore exponential in $\ell$
(we will see below that $M<1$ except at the critical point).

\subsection{Subcritical and critical cases}

Now we need to consider separately the three regimes.
First we deal with the subcritical phase, 
defined by
$\langle K \rangle < 1$, for which the fixed point is $q=1$ and then,
by the properties of the probability generating function, 
$M=f'(1)=\langle K \rangle$
and $2C=f''(1)=\langle K (K-1)\rangle= 
\sigma^2 +\langle K \rangle (\langle K \rangle -1)$,
where $\sigma^2 $ is the variance of $K$.
This leads to 
$$
 {c_{n+\ell}} 
=\frac {c_{n}} {\langle K \rangle ^\ell} +  
\frac{\sigma^2 (1-\langle K \rangle^\ell)}{2\langle K \rangle^{\ell+1}(1-\langle K \rangle)}-\frac{1-\langle K \rangle^\ell}{\langle K \rangle^\ell},
$$
where we will introduce the rescaled distance to the critical point
(distance in units of $1/\ell^{1/\nu}$), as
$$
y=\ell^{1/\nu}  (\langle K \rangle -1),
$$
this means that $\langle K \rangle=1+y/\ell^{1/\nu}$.
Further, we will take
$\ell \rightarrow \infty$, 
then, an interesting limit emerges for $\nu=1$,
yielding
$\langle K \rangle^\ell \rightarrow e^y$.
In order to keep $y$ finite we will impose
that we are in the vicinity of the critical point, 
$\langle K \rangle \rightarrow 1^-$.
Notice that in this limit only the middle term in the expression for 
${c_{n+\ell}}$ survives, 
i.e., 
$$
 {c_{n+\ell}} 
=-\frac{\sigma^2 (1-e^y) \ell}{2e^{y} y},
$$ 
and from here we can obtain the probability of surviving
in a system of limited size $L=n+\ell$, as
$$
P_\mathrm{surv}(L)=1-P_\mathrm{ext}(L)=1-f^L(0)=\frac 1 {c_L}=
\frac{2e^{y} y}{\sigma_{\hspace*{-0.5mm}\mathrm{c}}^2 (e^y-1) L},
$$
where we have considered $L \gg n$, and so $\ell \simeq L$,
and also have taken the variance right at the critical point, 
$\sigma_{\hspace*{-0.5mm}\mathrm{c}}^2$.

One can realize that this result also includes the critical case, 
given by $M=\langle K \rangle =1$, just taking the limit $y\rightarrow 0$,
for which $2 y e^y/(e^y-1) \rightarrow 2$. 
This is in correspondence with the replacement of Eq. (\ref{dearriba})
by $c_{n+\ell}=c_n + C\ell$. 
Therefore,
$$
P_\mathrm{surv}(L)=
\frac{2}{\sigma_{\hspace*{-0.5mm}\mathrm{c}}^2  L},
$$
at the critical point. This was apparently first proved
by Kolmogorov under more restrictive assumptions \cite[p. 20]{Harris_original},
\cite[p. 19]{Athreya}, \cite[p. 47]{branching_biology}.

\subsection{Supercritical case}

We show now that the supercritical case leads, 
through a more involving path, to exactly the same result
as the subcritical case. 
The main difference is that $M$ is not anymore the derivative
of $f(z)$ at $z=1$ but at $z=q<1$.
Let us expand by Taylor $f(q)$  
around $q=1$,
i.e., close to the value of the fixed point corresponding to the critical point,
$$
f(q)=1+\langle K \rangle (q-1) +\frac{f''(1)} 2 (q-1)^2 =q,
$$
which leads, for $q\ne 1$, to 
\begin{equation}
q=1- \frac 2 {f''(1)} (\langle K \rangle -1).
\label{q}
\end{equation}
Substituting in the 
Taylor expansion of $f'(q)$, 
up to first order in $q-1$,
$$
M=f'(q) = \langle K \rangle +f''(1) (q-1)=2-\langle K \rangle,
$$
which is smaller than 1 as $ \langle K \rangle >1$, 
and using that $\langle K \rangle=1+y/\ell$,
allows one to calculate $M^\ell=e^{-y}$ and $1-M=y/\ell$.
Notice that this looks the same as in the subcritical case
but replacing $y$ by $-y$.
Going back to Eq. (\ref{dearriba}),
$$
 {c_{n+\ell}} 
=\frac{\sigma^2 (1-e^{-y}) \ell}{2e^{-y} y},
$$
when $\ell$ is very large, approximating $2C=f''(q)=f''(1)=\sigma^2+\langle K \rangle (\langle K \rangle-1)$, and therefore 
$$
q-f^L(0) = \frac 1 {c_L} = \frac{2 y e^{-y}}{\sigma_{\hspace*{-0.5mm}\mathrm{c}}^2 (1-e^{-y}) L},
$$
introducing the variance at the critical point, $\sigma_{\hspace*{-0.5mm}\mathrm{c}}^2$.

In order to obtain the probability of surviving,
we need to add $1-q$, which goes, using Eq. (\ref{q}), as
\begin{alignat*}{1}
1-q&= \frac{\langle K \rangle -1}{f''(1)/2}=
2 \frac{\langle K \rangle -1}{\sigma_{\hspace*{-0.5mm}\mathrm{c}}^2}\left(1- 
\frac{\langle K \rangle(\langle K \rangle-1)}{\sigma_{\hspace*{-0.5mm}\mathrm{c}}^2}\right)=
\\
&=\frac{2y}{\sigma_{\hspace*{-0.5mm}\mathrm{c}}^2 L}\left(1-\frac y {\sigma_{\hspace*{-0.5mm}\mathrm{c}}^2 L} -\frac {y^2}
{ \sigma_{\hspace*{-0.5mm}\mathrm{c}}^2 L^2} \right),
\end{alignat*}
using also $\langle K \rangle-1=y/L$.
Therefore, the leading term in $P_\mathrm{surv}(L)$ turns out to be
\begin{alignat*}{1}
P_\mathrm{surv}(L)&=1-f^L(0) =1-q + \frac 1 {c_L} =
\\
&=\frac 1 {\sigma_{\hspace*{-0.5mm}\mathrm{c}}^2 L}\left(
{2y}+\frac{2e^{-y} y}{1-e^{-y}}\right)
=\frac 1 {\sigma_{\hspace*{-0.5mm}\mathrm{c}}^2 L}\left( \frac{2ye^y}{e^y-1} \right)
\end{alignat*}
which is the same indeed as in the subcritical case.

\subsection{Finite-size scaling law}

\begin{figure}
\subfigure[\  \label{fig1a}]{\includegraphics[width=0.48\textwidth]{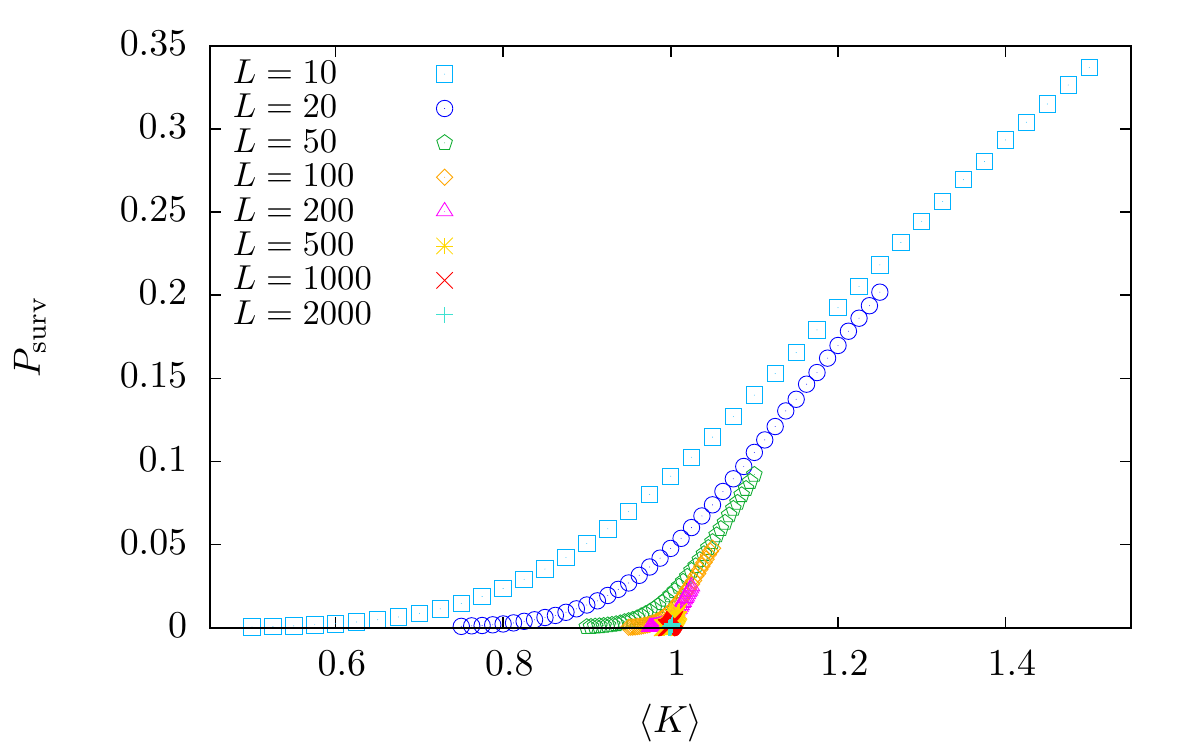}}
\subfigure[\  \label{fig1b}]{\includegraphics[width=0.48\textwidth]{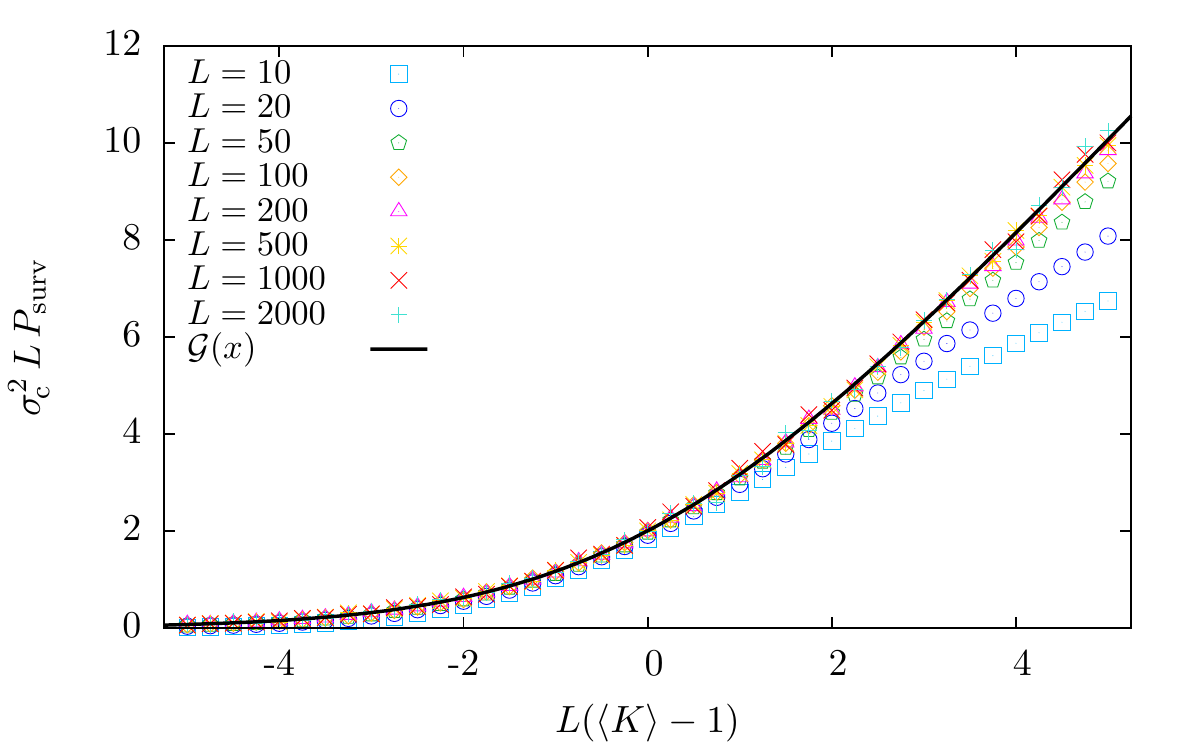}}
\caption{\label{fig1}
Validity of the finite-size scaling law for the geometric Galton-Watson process.
\subref{fig1a}
Simulation results for the 
probability of surviving as a function of the mean number
of offsprings $\langle K \rangle$ in a Galton-Watson process defined
by a geometric distribution of $K$, given by Prob$[K=k]=p (1-p)^k$,
for $k=0,1, \dots$
Different values of $L$ show the dependence with system size.  
Probabilities are estimated from 1000 independent realizations.
\subref{fig1b} Same probabilities as a function of $\langle K \rangle-1$
under rescaling with $L$ and $\sigma_{\hspace*{-0.5mm}\mathrm{c}}$. 
The collapse of the curves onto a unique scaling function, given by our theoretical result, is the signature of the validity
of the finite-size scaling for large system sizes $L$ and 
close to the critical point $\langle K \rangle=1$. 
}
\end{figure}

\begin{figure}
\includegraphics[width=0.48\textwidth]{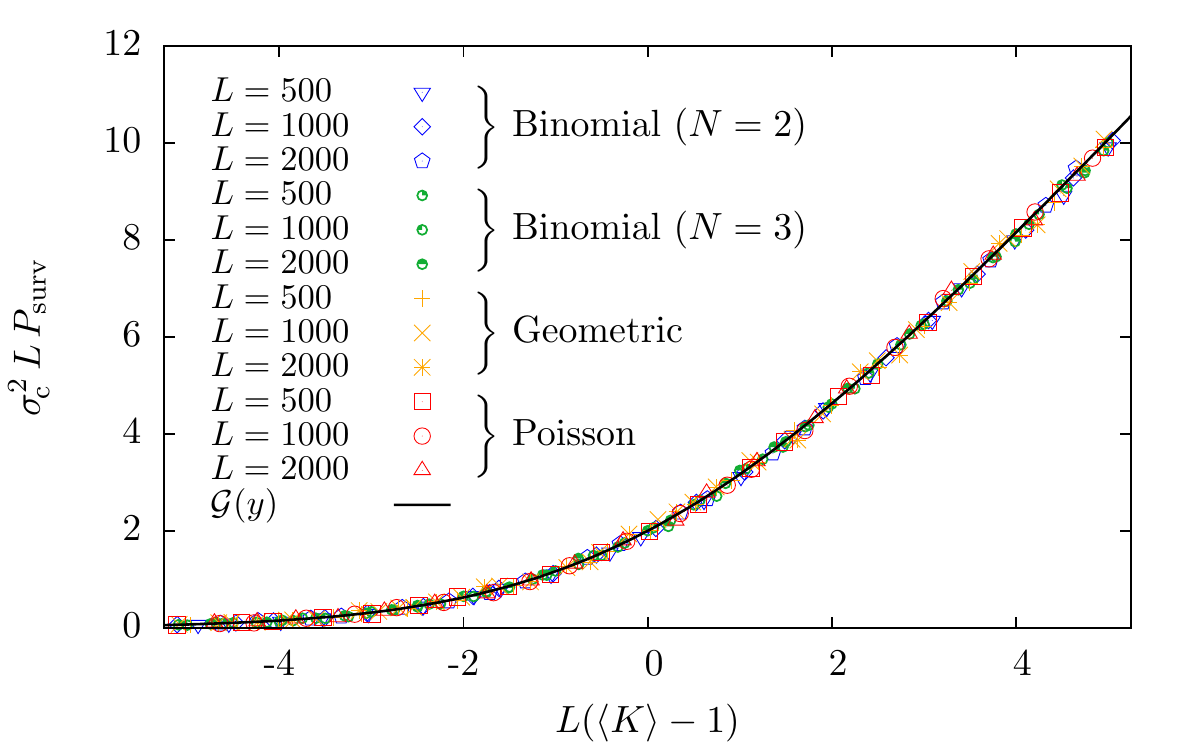}
\caption{
Universality of the finite-size scaling.
Extension of Fig.~\ref{fig1}(b) to other Galton-Watson processes
with different distributions for the number of offsprings $K$.
In addition to the geometric case,
the binomial distribution, with Prob$[K=k]=N ! p^k (1-p)^{N-k}/(k!(N-k)!)$
and $N=2$ or $3$,
and the Poisson distribution, with Prob$[K=k]=e^{-\lambda} \lambda^k /k!$, are simulated for different $\langle K\rangle$ and 
different system sizes $L$.
The collapse of all rescaled curves validates the universality of the scaling exponents and of our scaling function.
}
\label{fig2}
\end{figure}

The previous formula, and its identical replication in the subcritical and critical cases, allows
one to write the relationship between
the probability of surviving and the control parameter $\langle K \rangle$
in the form of a scaling law,
\begin{equation}
P_\mathrm{surv}(L)=\frac1 {L \sigma_{\hspace*{-0.5mm}\mathrm{c}}^2} { \mathcal{G}(L(\langle K \rangle-1))},
\label{Psurv_scaling}
\end{equation}
with  
$$
\mathcal{G}(y)=\frac {2 y e^y}{e^y-1}
$$ 
a universal scaling function, as it is independent of the underlying distribution
of $K$ (number of offsprings per element). 
This is valid for large system sizes $L$ and
small distances to the critical point, keeping $L(\langle K \rangle-1)=y$ finite.
Notice that this corresponds to the usual finite-size scaling form, Eq. (\ref{ffs_original}),
at zero field, $h=0$, with exponents $$\beta=\nu=1$$ (and metric factor $b=1$).
The only peculiarity is the appearance of an additional metric factor
given by $\sigma_{\hspace*{-0.5mm}\mathrm{c}}^2$.
The validity of the finite-size scaling law and its universality is confronted with computer simulations
of diverse branching processes in Figs. \ref{fig1} and \ref{fig2},
with positive results.

From the finite-size scaling law, it is remarkable that
keeping fixed the rescaled distance to the critical point, $y$,
the survival probability decays to zero with $L$ as a power law, 
hyperbolically, see Eq. (\ref{Psurv_scaling}).
In contrast, when the absolute distance to the critical point,
$\langle K 	\rangle-1$, is fixed, the decay is exponential towards the
infinite-size probability, 
$P_\mathrm{surv}(L)=1-q +[f'(q)]^L$, 
see Refs. \cite[p. 16]{Harris_original} and
\cite[p. 38]{Athreya}, or Eq. (\ref{dearriba}) also.

Interesting information comes from the different limit behaviors
of the scaling function,
$$
\mathcal{G}(y) \rightarrow \left\{
\begin{array}{rl}
-2ye^y & \mbox{when } y\rightarrow -\infty,\\
2 & \mbox{when } y\rightarrow 0,\\ 
2y & \mbox{when } y\rightarrow \infty,\\
\end{array}
\right.
$$
which yields,
$$
P_\mathrm{surv}(L) \rightarrow \left\{
\begin{array}{ll}
2\sigma_{\hspace*{-0.5mm}\mathrm{c}}^{-2}(1-\langle K \rangle) e^{-L(1-\langle K \rangle)} &\mbox{for } \langle K \rangle < 1, \\ 
2\sigma_{\hspace*{-0.5mm}\mathrm{c}}^{-2} L^{-1} &\mbox{for } \langle K \rangle=1, \\                
2(\langle K \rangle-1)/\sigma_{\hspace*{-0.5mm}\mathrm{c}}^2 &\mbox{for } \langle K \rangle >1.\\ 
\end{array}
\right.
$$
The subcritical and supercritical results have to be understood as 
holding close to the critical point but for an infinitely large system.

\section{Discussion}

We can particularize the finite-size scaling law,
Eq. (\ref{Psurv_scaling}), to the case where
the number of offsprings $K$ of each element is given
by a binomial distribution,
defined by $N$ trials with probability of success $p$ in each trial.
This distribution is relevant for the analogy with percolation
\cite{Christensen_Moloney}.
The mean and variance are given by $\langle K\rangle = Np$
and $\sigma^2=Np(1-p)$. The critical point arises then at
$p_c=1/N$ and $\sigma_{\hspace*{-0.5mm}\mathrm{c}}^2=1-1/N$, and the finite-size scaling
can be written as
$$
P_\mathrm{surv}(L)=\frac 1 {L(1-1/N)} \mathcal{G} \left(NL\left(p-\frac 1 N\right)\right),
$$
where the universality of Eq. (\ref{Psurv_scaling}) becomes ``diffused''
due to the appearance of two non-universal metric factors multiplying $L$.
But it is worth mentioning that $P_\mathrm{surv}$ is not the
order parameter defined usually in percolation, see for instance
Ref. \cite{Botet}.

As another illustration, if we considered a geometric distribution 
for $K$ (defined for $K\ge 0$), with success parameter $p$,
the mean and variance are $\langle K\rangle =(1-p)/p $
and $\sigma^2=(1-p)/p^2$. The critical point is then $p_c=1/2$, 
$\sigma_{\hspace*{-0.5mm}\mathrm{c}}^2=2$ and then the finite-size scaling transforms into
$$
P_\mathrm{surv}(L)=\frac 1 {2 L} \mathcal{G} \left(-4 L\left(p-\frac 1 2\right)\right).
$$
Notice that in this case the parameter $p$ has a different interpretation
as for the binomial distribution.
In any case, Fig. \ref{fig2} demonstrates that the same universal finite-size scaling
holds for the binomial, the geometric, and the Poisson distributions.

Summarizing, 
we have found that the second-order phase transition 
from sure extinction to non-sure extinction in the Galton-Watson
branching process fulfills a finite-size scaling law, where the scaling
function and the scaling exponents can be exactly derived.
If the variance of the distribution of the number of offsprings per
element is taken into account in the scaling law, this becomes universal,
with universal metric factor, in the sense that it is exactly the same for
all Galton-Watson branching processes.

\section*{Acknowledgements}

R. Garcia-Millan has enjoyed from a stay at the Centre de Recerca Matem\`atica
through its Internship Program.
The rest of authors acknowledge support from projects
FIS2012-31324, from Spanish MINECO, and 2014SGR-1307, from AGAUR.
F. Font-Clos is also grateful to the AGAUR for a FI grant.

\bibliographystyle{unsrt}

\bibliography{../../../../projects/words_ramon/p1_lemmas/biblio}

\end{document}